\documentclass[twocolumn,1pt]{revtex4-1}
 \usepackage{amsmath,amssymb}
 \usepackage{bm}
 \usepackage[dvipdfmx]{graphicx}
\usepackage[top=23truemm,bottom=23truemm,left=15truemm,right=15truemm]{geometry}
\usepackage{color}
\usepackage{here}
\begin{document}
 \title{
 Exponential system-size dependence of the lifetime of transient spiral 
 chaos \\in excitable and oscillatory media
 }
 \author{Kaori Sugimura}
 \email[corresponding author: ]{sugimura.kaori@is.ocha.ac.jp}
\author{Hiroshi Kori}
 \affiliation{Department of Information Sciences, Ochanomizu University, Tokyo 112-8610, Japan}
 \date{\today}

 \begin{abstract}
  Excitable media can develop spiral chaos, in which the number of
  spirals changes chaotically with time. Depending on parameter values in
  dynamical equations, spiral chaos may permanently persist or
  spontaneously arrive at a steady state after a transient time, referred
  to as the lifetime. Previous numerical studies have demonstrated that
  the lifetime of transient spiral chaos increases exponentially with
  system size to a good approximation.  In this study, using the fact
  that the number of spirals obeys a Gaussian distribution, we provide a
  general expression for the system size dependence of the lifetime for
  large system sizes, which is indeed exponential.
 We confirm that the expression is in good agreement with
  numerically obtained lifetimes for
  both excitable and oscillatory media with parameter sets near the onset of
  transient chaos.
  The expression we develop for the lifetime is expected to be
  useful for predicting
  lifetimes in large systems.
 \end{abstract}

 \maketitle
 \section{INTRODUCTION}
 Excitable media play vital roles in various systems \cite{winfree80,keener09,cross93}.
 Excitable media in biological tissues support the
 propagation of signals, such as concentration waves in the heart and
 electrical impulses in nerve axons.
 Such waves are also used for communication between certain microorganisms ({\it{Dictyostelium discoideum}}).

 Moreover, excitable media exhibit a particular type of
 spatiotemporal chaotic dynamics, in which
 spiral waves spontaneously generate or annihilate (spiral
 chaos) \cite{winfree80,kuramoto84}.
 Spiral chaos is commonly observed in surface reaction systems
 \cite{beta06, baer93}.
 Similar chaotic dynamics are also observed in the heart, causing
 fibrillation \cite{qu06}.
 So far, several mathematical models for excitable media that exhibit 
 spiral chaos have been proposed \cite{baer93,qu06, krefting10}.

It is also known that spiral chaos may develop in oscillatory media,
 e.g., those obeying the complex Ginzburg-Landau equation (CGLE) \cite{kuramoto84, chate96, krefting10, berenstein11}.
 In such mathematical models,
 depending on the parameter values, spiral chaos permanently persists or
 spontaneously terminates
 (Fig.~\ref{fig:spiral_chaos}).
  In the latter case, the system eventually arrives at a steady state after a transient time,
 which we refer to as a lifetime. 
 The dependence of the lifetime of spiral chaos on the system size
 has received much attention
 in the context of the clinical treatment of cardiac fibrillation
 (\cite{qu06} and the references therein).
 In \cite{qu06}, it is numerically demonstrated using
 both a variant of FitzHugh--Nagumo model (referred to as the B\"{a}r
 model \cite{baer93}) and a more realistic model for cardiac
 electrical dynamics that the lifetime increases exponentially with the system size.
 Such an exponential dependence, as well as hyperexponential dependences, had
 already been reported in other types of
 transient chaos \cite{wackerbauer03,tel08,wacker95,kaneko88}.

 The main focus of the present study is an expression for the
 dependence of the lifetime of spiral chaos in excitable media on the
 system size.
 For this goal, we 
 first investigate statistical properties regarding
 the number of spiral cores (namely, defects).
 There is a large body of studies on such statistical properties
 \cite{gil90,ebata11,qiao09,davidsen03,falcke99}.
 In particular, it is known that as system size increases,
 the probability distribution of the number of
 defects during transient spiral chaos approaches a Gaussian
 distribution \cite{davidsen05}, as is naturally expected from
 the central limit theorem.
 Using this fact,
 we derive an expression for the
 system size dependence of the lifetime, which is indeed exponential.

 We extensively investigate the system size dependence of the lifetime
 using two different models, the B\"{a}r model and the CGLE,
 with several parameter sets and different boundary conditions.
 We find that while the lifetime increases
 exponentially with system size in all cases, our expression fits well
 for parameter sets near the onset of transient chaos,
 suggesting that some assumptions may be violated depending on parameter values.

 The present paper is organized as follows. In Sec.~\ref{sec:model}, we
 describe the model and the numerical settings. In
 Sec.~\ref{sec:distribution},
 we show that the probability distribution of the number of
 defects approaches a Gaussian distribution as the system size increases.
 In Sec.~\ref{sec:lifetime}, we first numerically show that the lifetime of transient
 spiral chaos increases exponentially with system size; then, we
 derive the expression for the system size dependence of the lifetime,
 which fits well to numerical data for some parameter sets.
 A summary and discussion are provided in Sec.~\ref{sec:conclusions}.

 \section{model and numerical settings}\label{sec:model}
\begin{figure*}[t]
\begin{minipage}{0.25\hsize}
\begin{center}
 \includegraphics[scale=1.0]{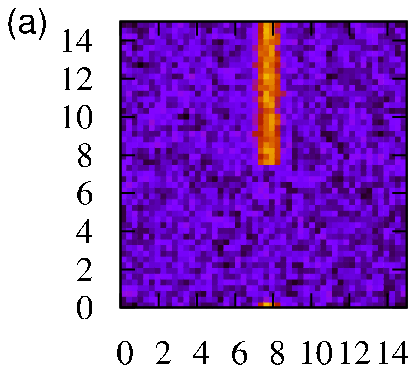}
\end{center}
 \end{minipage}
\begin{minipage}{0.25\hsize}
\begin{center}
 \includegraphics[scale=1.0]{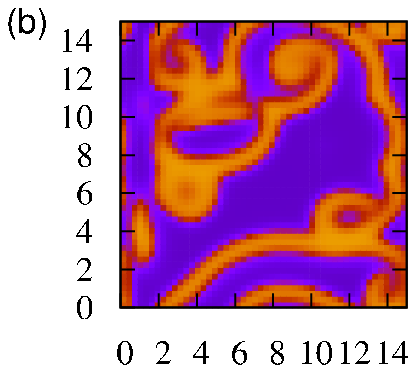}
\end{center}
 \end{minipage}
\begin{minipage}{0.25\hsize}
\begin{center}
 \includegraphics[scale=1.0]{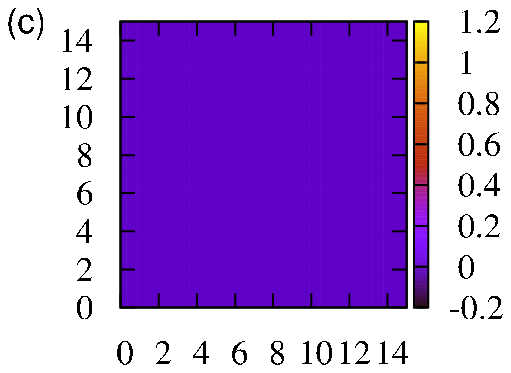}
\end{center}
 \end{minipage}
 \caption{Snapshots of $u(x,y,t)$ in the B\"{a}r model.
(a) Initial condition ($t=0$) is constructed as follows. First, we create a flat
 excitation wave, run a simulation for a while, and
 cut the wave into half. We then
 add random noise with a uniform probability distribution
 over $[-0.25,0.25]$.
(b) Transient spiral chaos ($t=136$).
(c) Uniform steady state ($t=160$).
 The parameter values are
 $a=0.84,~b=0.07,~\epsilon=0.08$, and $N=15^2$.
}

 \label{fig:spiral_chaos}
\end{figure*}

 For most of our numerical investigation, we employ the B\"{a}r model \cite{baer93},
 which is a modified FitzHugh--Nagumo model representing an excitable medium.
 This model has also been employed in \cite{qu06}.
 The model gives
 \begin{subequations}
  \label{baer_model}
  \begin{eqnarray}
   \frac{\partial{u}}{\partial{t}} &=& -\frac{1}{\epsilon}u(u-1)\left(u-\frac{v+b}{a}\right)+D\nabla^2u,\\
   \frac{\partial{v}}{\partial{t}} &=& f(u)-v, \\
   f(u)&=&
    \begin{cases}
     0 \ , \ u<\frac{1}{3},\\
     1-6.75u(u-1)^2 \ , \ \frac{1}{3} \leq u \leq 1,  \\
     1 \ , \ u>1,
    \end{cases}
  \end{eqnarray}
 \end{subequations}
 where the parameters $\epsilon, a, b$ and the diffusion coefficient
 $D$ are positive.
 The system is two dimensional with an area $L \times L \equiv N$.
 The variables $u(x,y,t)$ and $v(x,y,t)$ are
 interpreted in the context of cell physiology
 as the membrane potential and the recovery variable, respectively
 \cite{keener09}.
Numerical simulations are performed using 
 the fourth-order Runge--Kutta method with space step $h = 0.3$
 and time step $s=0.01$.

 We first note that in Eq.~\eqref{baer_model}, the uniform steady
 state $u(x,y)=v(x,y)=0$
 is linearly stable for any set of parameter values.
 By inserting the ansatz $u, v \sim e^{\lambda t - {\rm i} \bm q \cdot
 \bm r}$ with $\bm r = (x, y)$ and a wave vector $\bm q$ of 
 perturbation, we obtain $\lambda = -1, -\frac{b}{a \epsilon}-D|\bm
 q|^2$, which is always negative.
 Hence, the system should smoothly arrive at the uniform steady
 state if the initial state is close to it.

 However, for appropriate parameter values and initial conditions,
 spatiotemporal chaotic dynamics arise (referred to as
 spiral chaos) (Fig. \ref{fig:spiral_chaos}).
 As reported in \cite{baer93}, for a broad range of $b$ ($b<0.18, a=0.84$), the
 following behavior arises. For small $\epsilon$ values ($0.01 <
 \epsilon < 0.06$), spiral waves rigidly rotate. For $\epsilon > 0.06$,
 spiral waves begin to meander. For $\epsilon > 0.07$, spiral chaos arises.
 In this region, spirals begin to
 break up after some
 transient rotations, resulting in the formation of two free ends of a wave.
 From these free ends, a new pair of counter-rotating spirals
 arise. There is also a pair-annihilation process, in which the cores of a pair of
 counter-rotating spirals collide and annihilate.
 Moreover, in the Neumann boundary condition, there is an additional case in
 which a defect is absorbed by the boundary.
 These processes are repeated chaotically.

 As a convenient initial condition for realizing this chaotic state,
 we employ a flat broken wave (Fig. \ref{fig:spiral_chaos}), in which
 there initially exists
 a defect
 for the Neumann boundary condition or a pair of defects 
 for the periodic boundary condition.
 To obtain statistically independent results for each run of the simulations,
 we add independent random noise
 obeying a uniform probability distribution over
 $[-\eta,\eta]$ with $\eta=0.25$ to $u$ and $v$ 
 at all discretized points at
 $t=0$.  Note that the evolution is noise free for $t>0$.
 In our preliminary numerical simulations, we
 have checked that our statistical results
 do no change quantitatively for $\eta=0.1$ (results not shown).
 The results presented assume the periodic boundary condition
 and $a=0.84, b=0.07, \epsilon=0.08, D=1$ unless otherwise noted.
 Some results are obtained with the Neumann boundary condition and/or 
 other sets of $b$ and $\epsilon$ values.

To check the generality of our argument, we also numerically investigate
 the oscillatory media described by the
 complex Ginzburg--Landau equation (CGLE), given by
 \begin{equation}
  \label{cgle}
  \frac{\partial W}{\partial t} = W + (1+{\rm i} c_1) \nabla^2 W - (1 + {\rm
   i} c_2) |W|^2 W,
 \end{equation}
 where $W(x,y,t) \in \mathbb C$ is the state variable and $c_1, c_2 \in \mathbb R$ are the
 parameters of this system \cite{kuramoto84}.

 \section{Time evolution and probability distribution of the number of
  defects}\label{sec:distribution}
\begin{figure}[b]
\begin{minipage}{\hsize}
\begin{center}
 \includegraphics[scale=0.4]{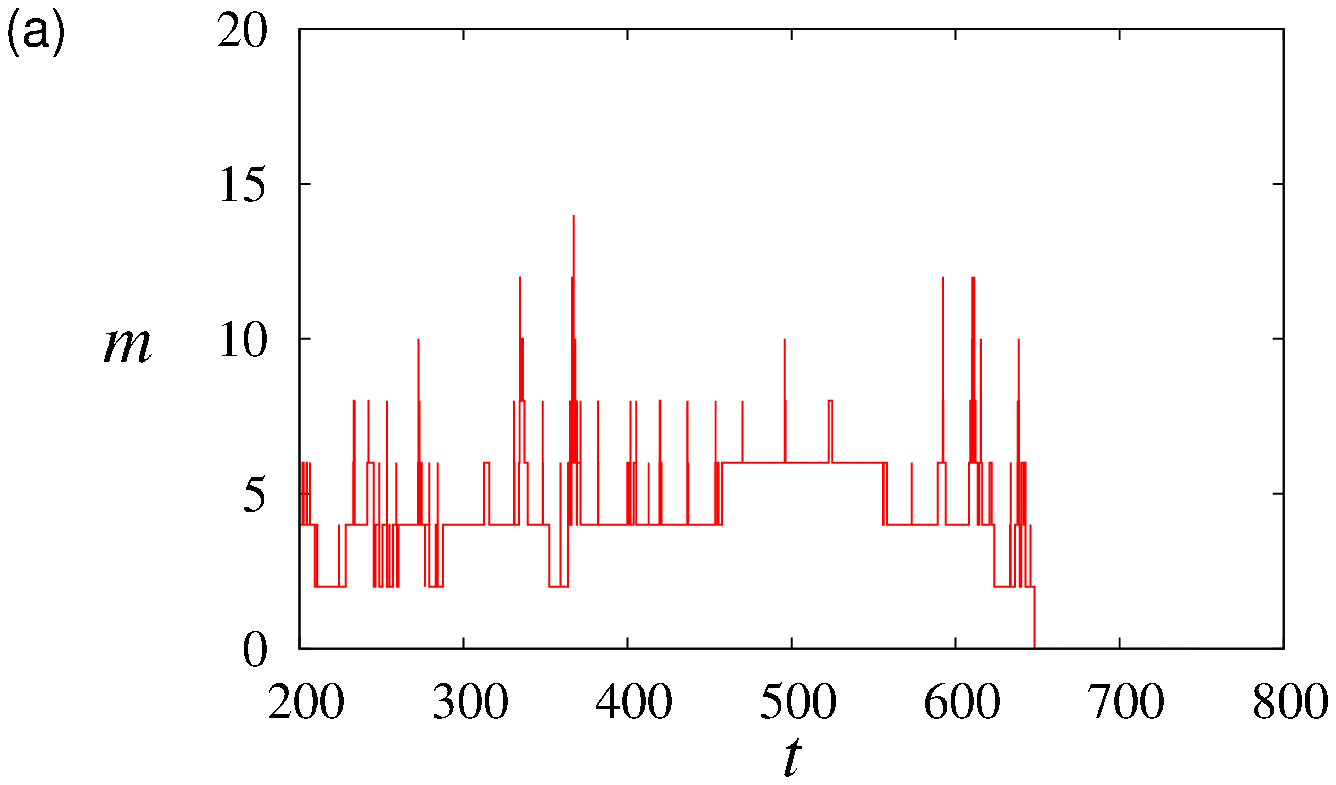}
\end{center}
 \end{minipage}\\
\begin{minipage}{\hsize}
\begin{center}
 \includegraphics[scale=0.4]{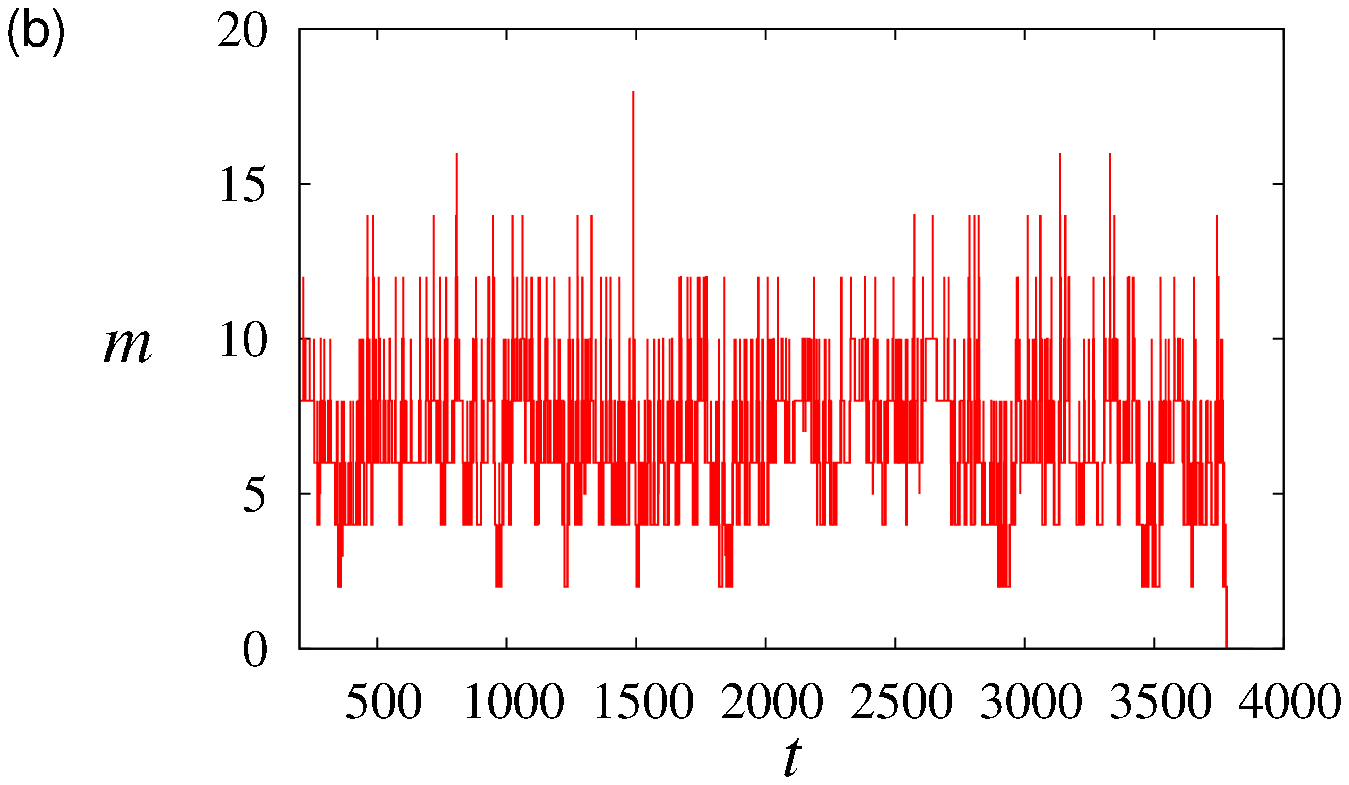}
\end{center}
 \end{minipage}
 \caption{Time series of the number $m(t)$ of defects with 
 system size (a) $N=24^2$ and (b) $N=30^2$. 
}
 \label{fig:time series of number of defects}
\end{figure}
\begin{figure*}[t]
\begin{minipage}{\hsize}
\begin{center}
 \includegraphics[scale=0.34]{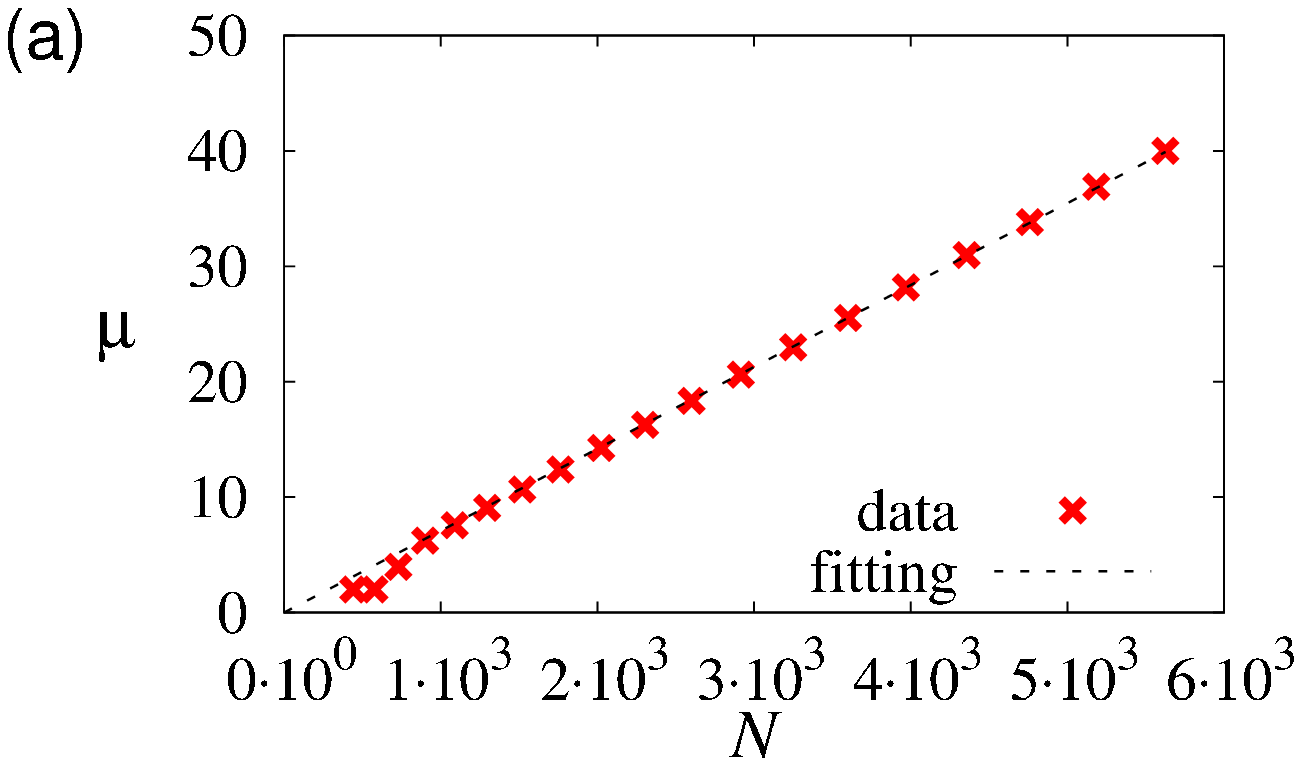}
 \includegraphics[scale=0.34]{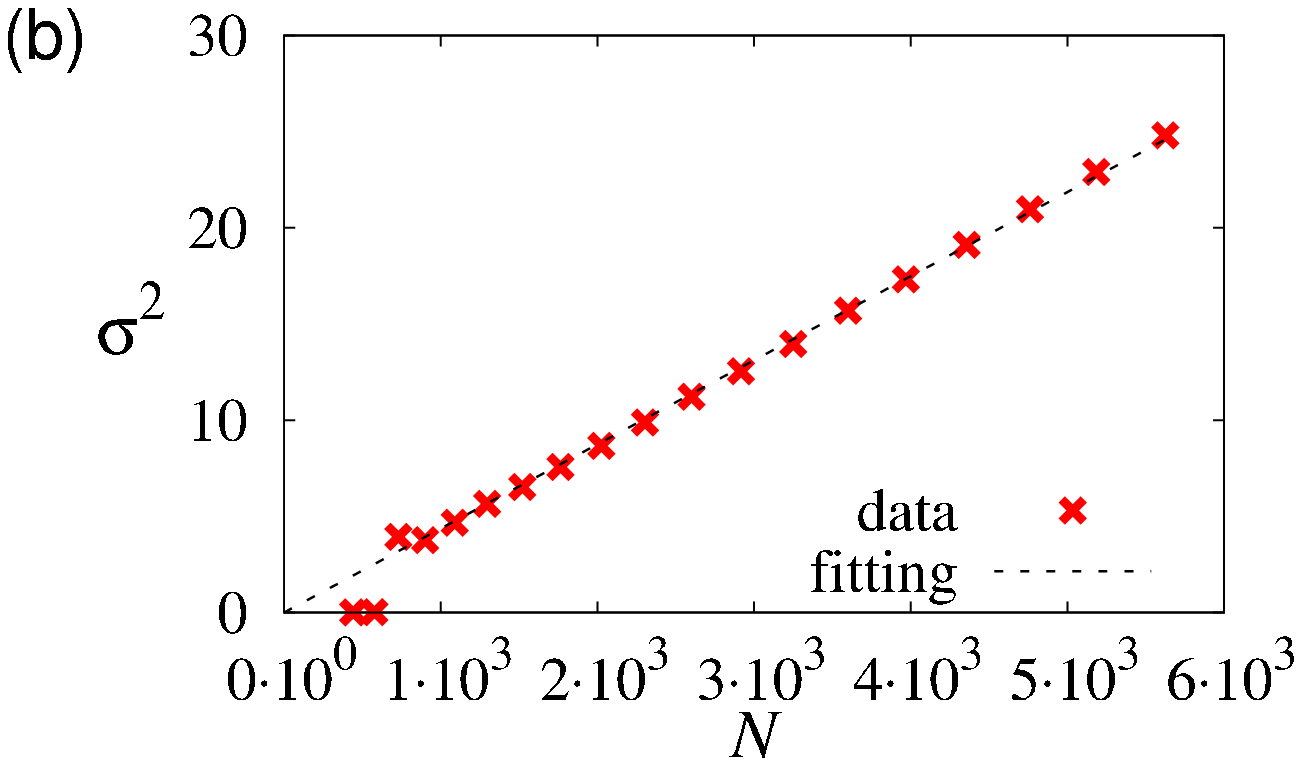}
\includegraphics[scale=0.34]{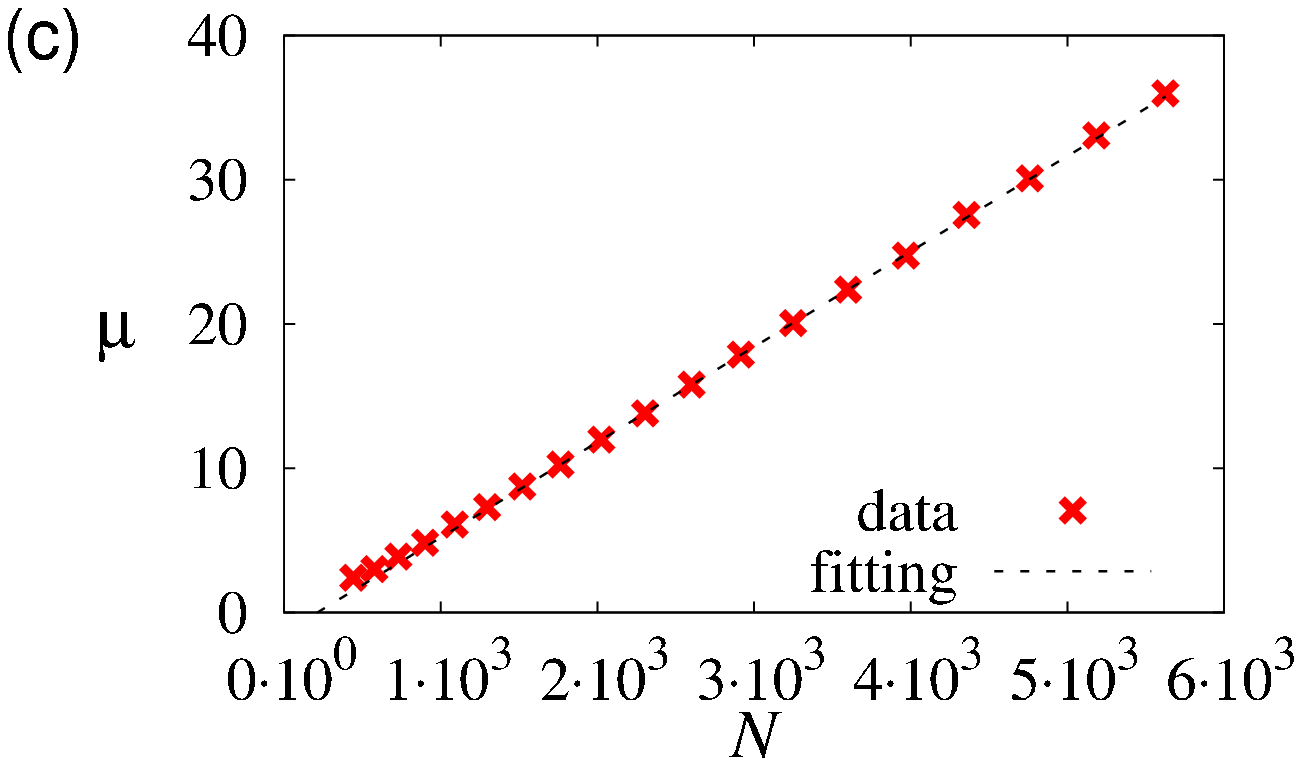}
\includegraphics[scale=0.34]{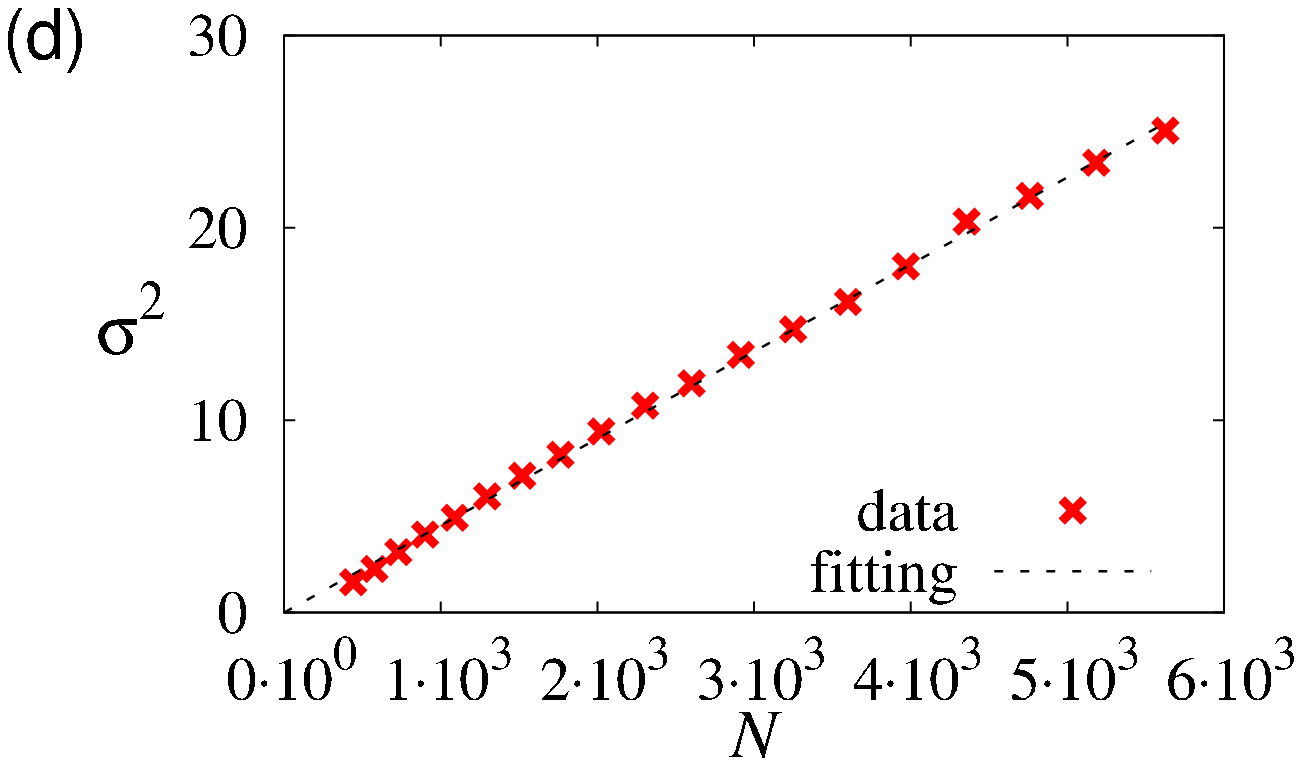}
\end{center}
 \end{minipage}
 \caption{Mean $\mu$ and variance $\sigma^2$ of the number of defects.
 (a, b) Results for the periodic boundary
 condition.
 The fitting lines are $\mu = 0.00709N$ and $\sigma^2  = 0.00437N$. (c, d) Results for the Neumann
 boundary condition. The fitting lines are  $\mu = 0.00660N-1.3996$
 and $\sigma^2 = 0.00452N$.
 Fitting is performed for data with $N>2000$.
 }
 \label{fig:mean}
\end{figure*}
 
We first investigate
 the time evolution and probability distribution of the number of
 defects. 
 All the results in this section are for periodic boundary
 condition.
 We confirmed that qualitatively the same
 results were obtained with the Neumann boundary condition.

The number $m(t)$ of defects at time $t$
 in the system was counted as follows.
 The phase $\phi(x,y)$ of the state is defined by
 $\arg [(u(x,y)-u_0) + {\rm i} (v(x,y)-v_0)]$ with $(u_0, v_0) = (0.5, 0.3)$
 and $\arg W$ for the B\"{a}r model and
 the CGLE, respectively.
 The topological charge $C(x,y,t)$ is defined by 
 $\frac{1}{2\pi}\oint \nabla \phi(\bm{r},t) \cdot d\bm{l}$.
 The defects with $C=1$ and $-1$
 are the cores of counterclockwise and clockwise spirals, respectively.
 The topological charge is numerically obtained by calculating
 $C(x,y)=(\phi_{1,2}+\phi_{2,3}+\phi_{3,4}+\phi_{4,1})/2\pi$, where $\phi_{i,j}=\phi_i-\phi_j$ ($-\pi\leq\phi_{i,j}<\pi$), $\phi_1 =\phi
 (x,~y)$, $\phi_2 = \phi(x+h,~y)$, $\phi_3 = \phi(x+h,~y+h)$, 
 $\phi_4=\phi(x,~y+h)$, and $h$ is the space step employed in our
 numerical simulations.
 We then reset $C=\pm 1$ when a numerically obtained $C$ value is in
 $[(\pm 2\pi-0.1)/2\pi,~(\pm 2\pi+0.1)/2\pi]$ and $C=0$ otherwise.
 The number $m(t)$ of defects is the sum of $|C|$ over the
 entire system.

 As seen in Fig. \ref{fig:time series of number of defects},
 $m(t)$ fluctuates strongly with time, and this chaotic process appears to be stationary.
 However, defects completely vanish at a certain time without any clear
 presage, and the
 system falls into the uniform steady state.
 As is the case in Figs.~\ref{fig:time series of number of defects} (a) and (b),
 a larger system
 typically has a larger number of defects and a longer transient time.

\begin{figure}[b]
\begin{minipage}{\hsize}
\begin{center}
 \includegraphics[scale=0.4]{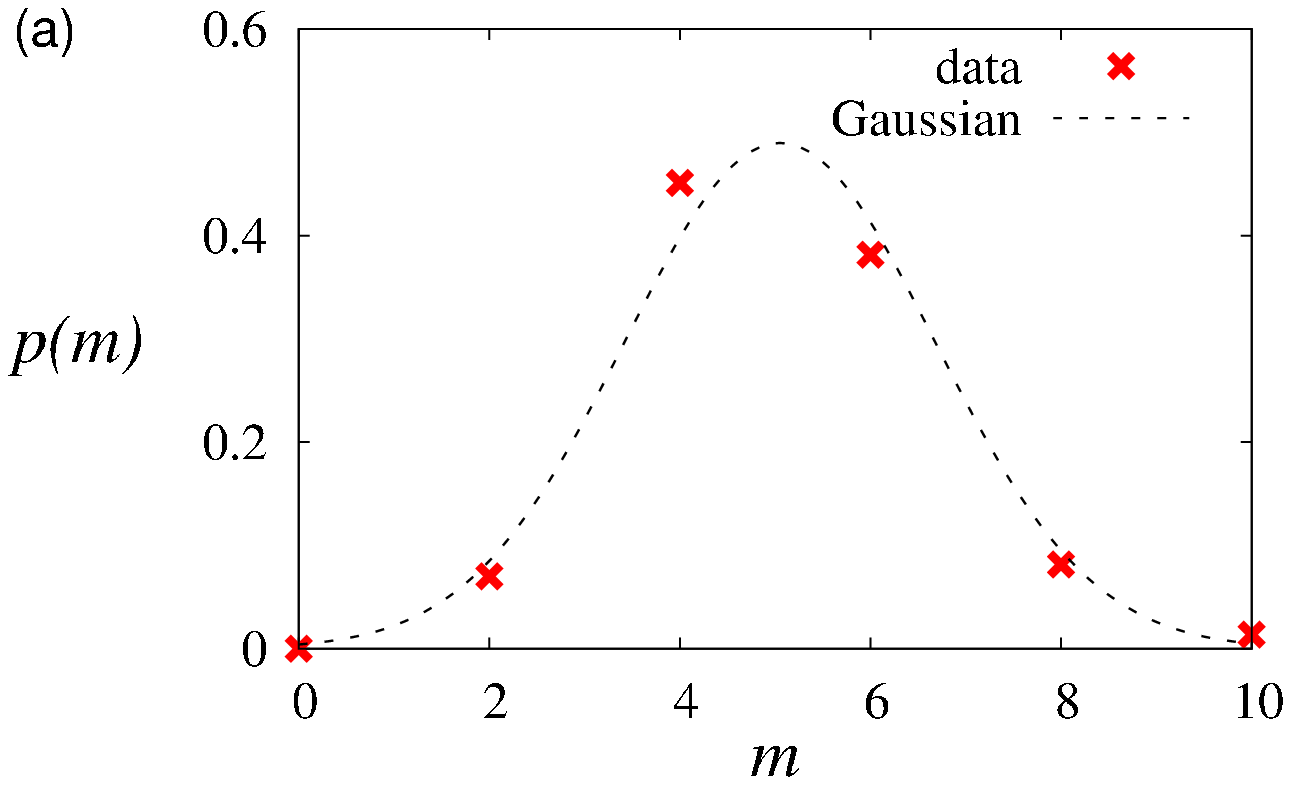}
\end{center}
 \end{minipage}
\begin{minipage}{\hsize}
\begin{center}
 \includegraphics[scale=0.4]{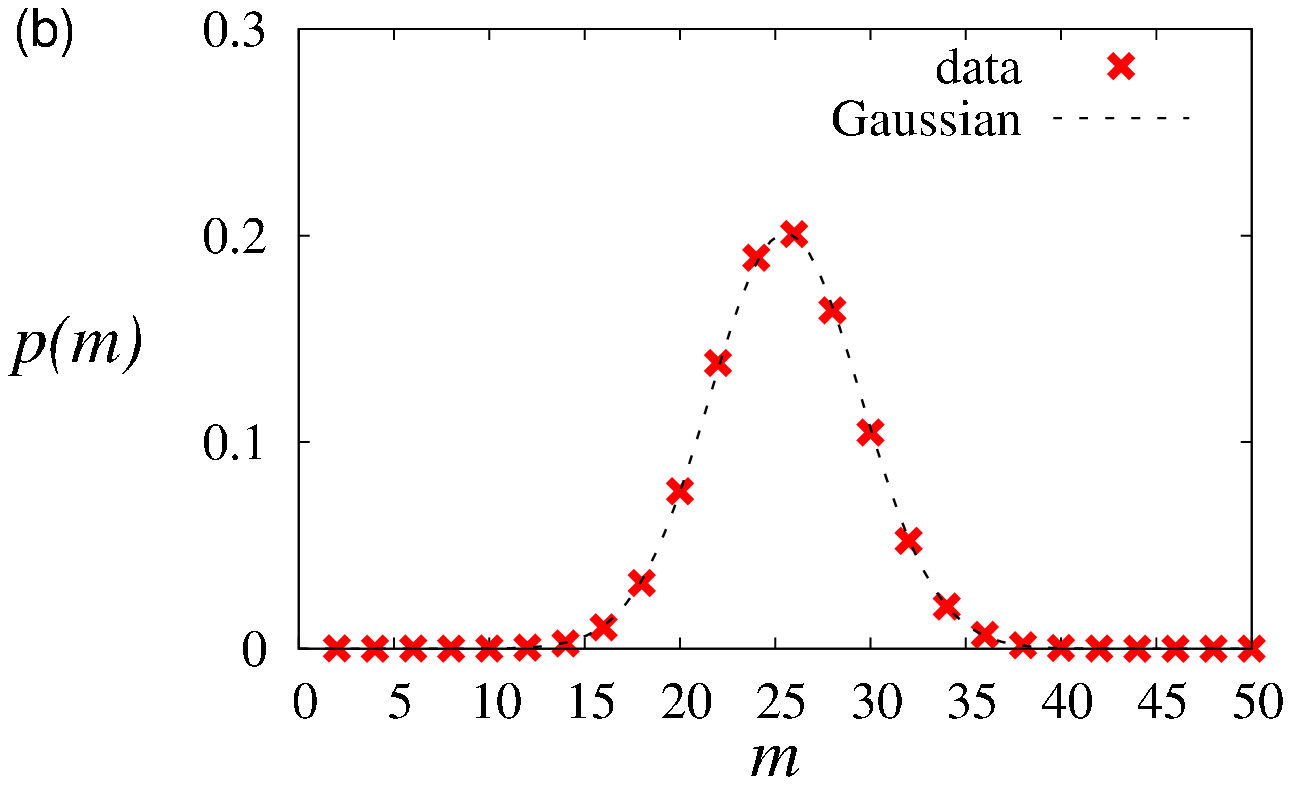}
\end{center}
 \end{minipage}

 \caption{Distribution of the number of defects. 
 (a) System size $N=27^2$. (b) $N=60^2$. 
 The dashed lines are the Gaussian distributions with average
 $\mu=\alpha N$ and variance $\sigma^2= \beta N$ with $\alpha=0.00709$
 and $\beta = 0.00437$.
 } 
 \label{fig:distribution}
\end{figure}

 Statistical properties are investigated with the time series of $m(t)$
 during transient chaos after the initial transient process ($t>100$)
 (Figs.~\ref{fig:mean} and \ref{fig:distribution}).
 Here for each system size, we employ many different initial conditions and
 the number of defects is counted at each time step 
 until the system arrives at the steady state.
 We find that both the mean $\mu$ and variance $\sigma^2$ 
 of $m(t)$ are approximately
 proportional to the system size $N$ (Fig.~\ref{fig:mean}):
 \begin{align}
  \mu & =\alpha N,\\
  \sigma^2 &= \beta N.
 \end{align}
 The linear growth of $\mu$ has also been found in \cite{strain98}.
 Next, we measure the probability distribution of the number of defects,
 which is the probability that there are $m$ defects at each time in the system during
 transient chaos.
 As is found in \cite{beta06}, we confirm that
 the probability distribution approaches
 the following Gaussian distribution
 as the system size increases (Fig.~\ref{fig:distribution}):
 \begin{align}
  p(m) &= \frac{\delta}{\sqrt{2\pi{\sigma}^2}}\exp
   \left[ -\frac{(m-\mu)^2}{2{\sigma}^2} \right]\\ 
 &= \frac{\delta}{\sqrt{2\pi{\beta}N}}\exp
   \left[ -\frac{(m-\alpha N)^2}{2{\beta}N} \right],
  \label{gaussian}
 \end{align}
 where
 $\delta=1$ for the Neumann boundary condition and $\delta=2$ for the
 periodic boundary condition because $m$ takes only
 even number values in the latter case.

\begin{figure}[t]
\begin{minipage}{\hsize}
\begin{center}
 \includegraphics[scale=0.38]{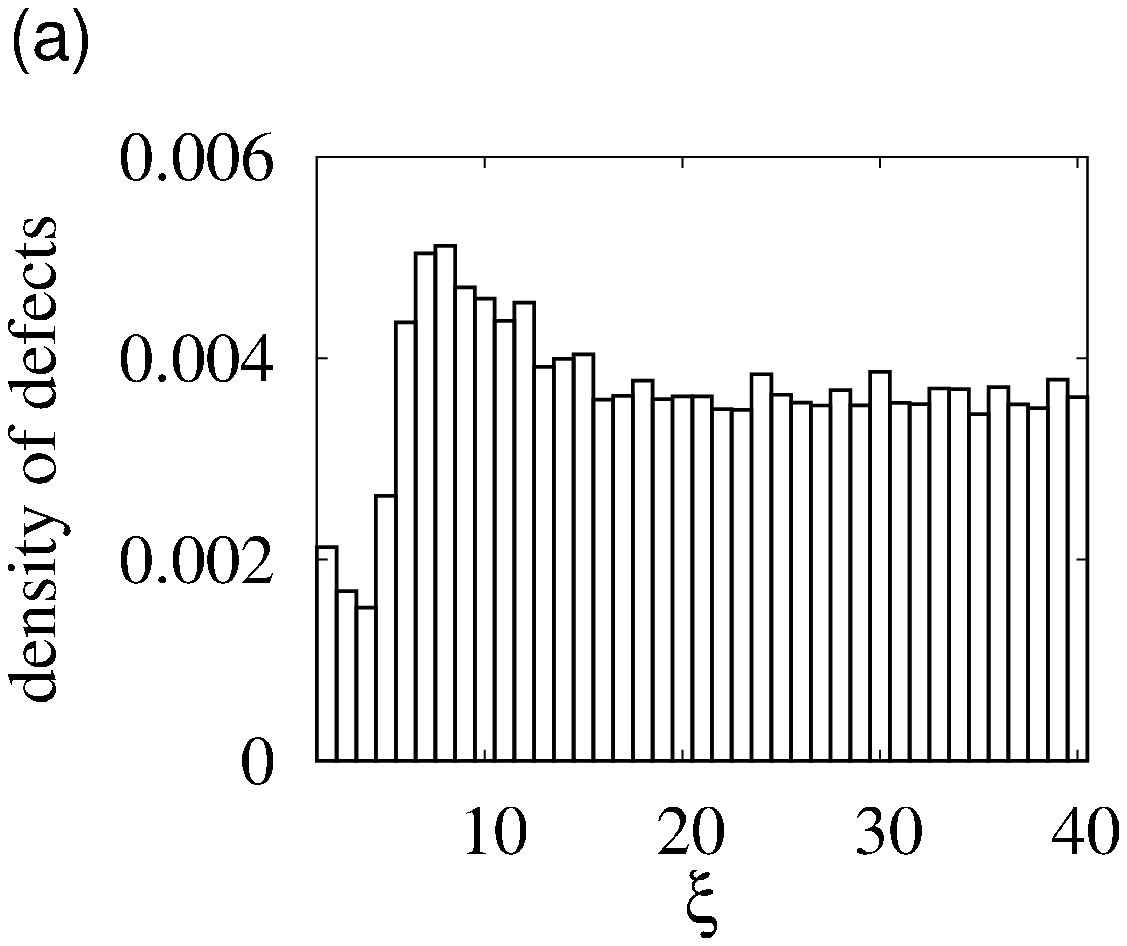}
 \includegraphics[scale=0.38]{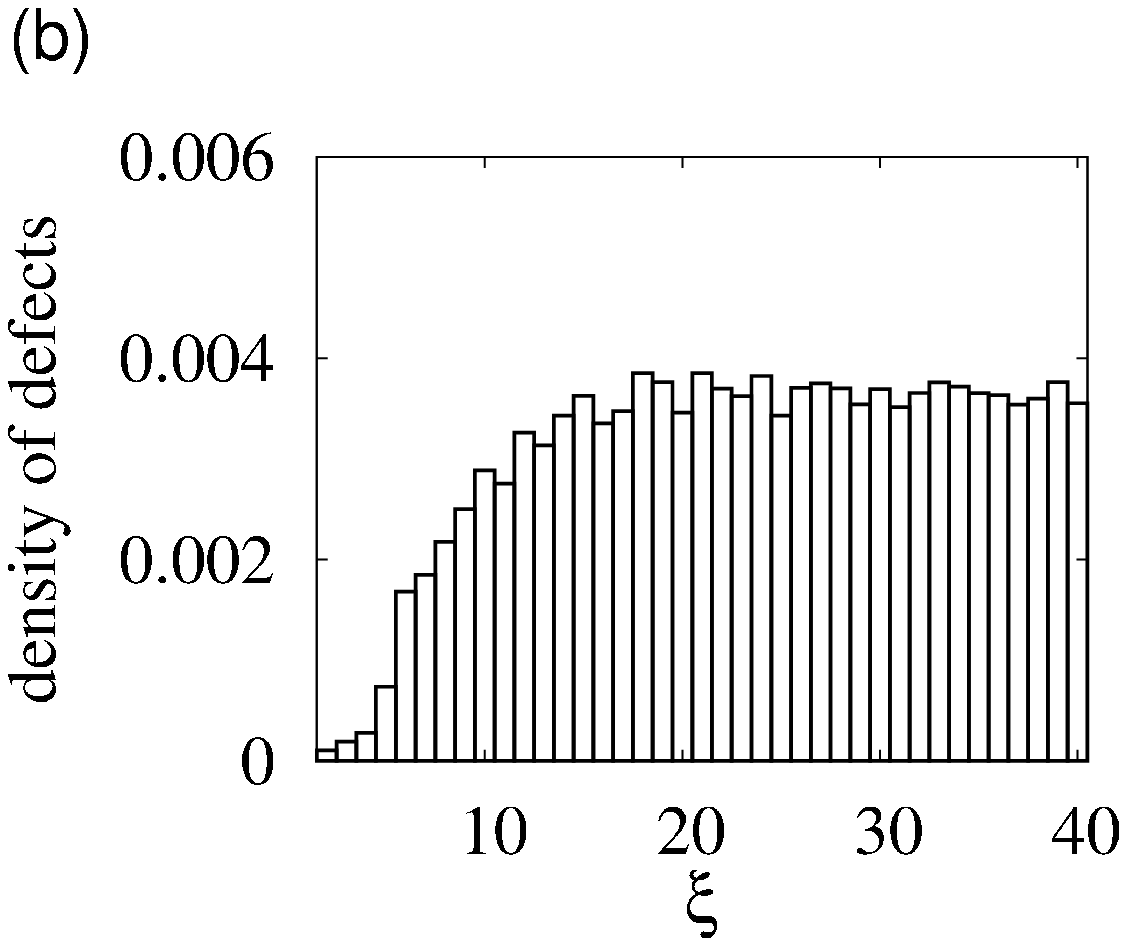}
 \includegraphics[scale=0.38]{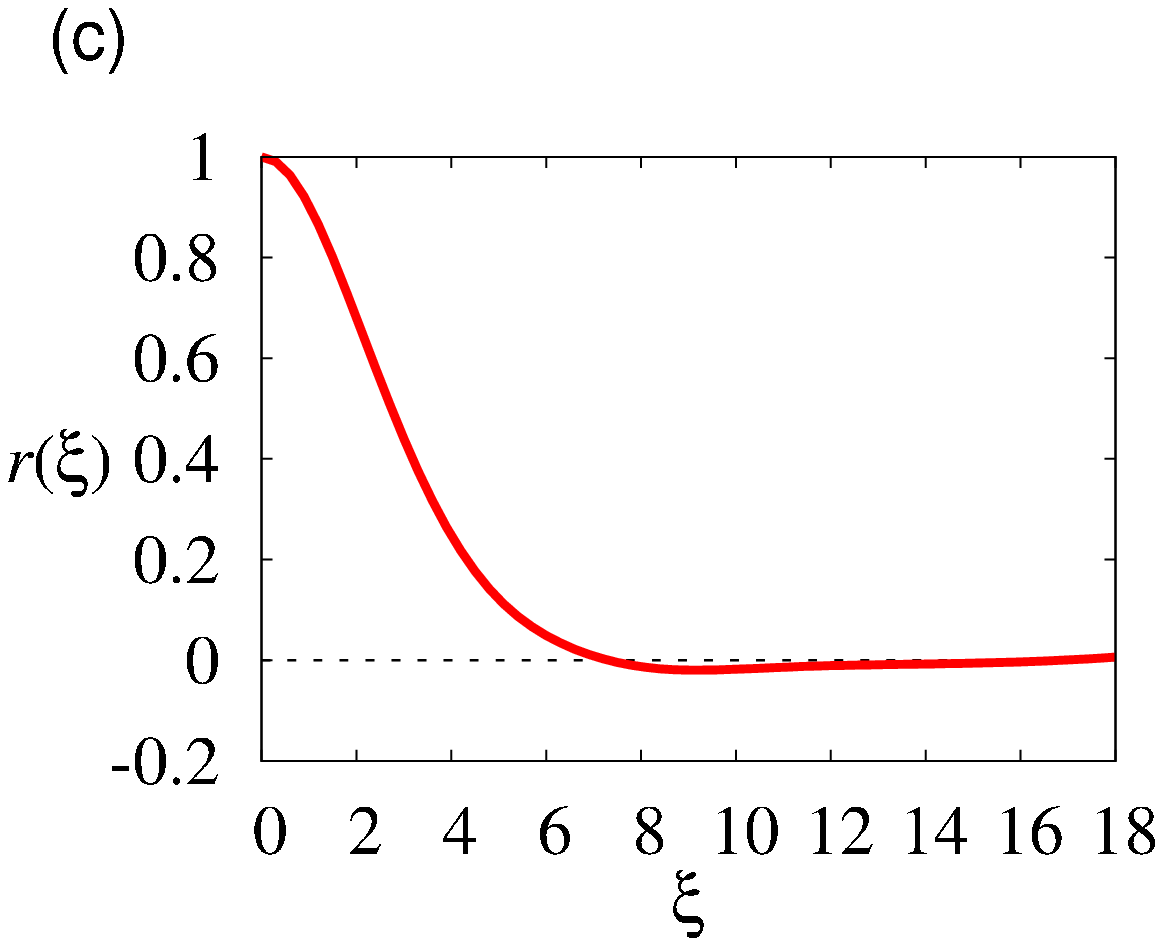}
\end{center}
 \end{minipage}

 \caption{Numerical measurement of correlation length.
 We estimate the correlation length using two methods. (a, b) The 
 density of defects with (a) $C=-1$ and (b) $C=1$ at $(x,y)$
 as a function of the distance $\xi=\sqrt{(x-x_0)^2+(y-y_0)^2}$ from a certain
 defect with $C=1$ at $(x_0,y_0)$ 
 (c) The Pearson product-moment
 correlation coefficient for the variable $u$, defined as
$
r(\xi)=\frac{
\int_0^T (u^*-\overline{u})(u-\overline{u})dt}{\sqrt{
\int_0^T (u^*-\overline{u})^2dt}\sqrt{
\int_0^T (u-\overline{u})^2dt}},
$
where $u^*=u(\frac{L}{2},~\frac{L}{2},~t)$,
$u=u(\frac{L}{2}-\xi,~\frac{L}{2},~t)$ and $\overline{u}$ is the average
 of $u$ over the entire system.
These results indicate that the correlation length is roughly $10$ or less.
 }
 \label{fig:correlation}
\end{figure}

 These results can be rationalized by the following argument.
 Suppose that the system is virtually divided into $n$ subsystems of
 size $\tilde L \times \tilde L = \tilde N$. 
 For the periodic boundary condition, all the subsystems should share
 a certain probability distribution of the number of
 defects with mean $\tilde \mu$ and variance ${\tilde
 \sigma}^2$.
 If the linear length $\tilde L$ of each subsystem is 
 sufficiently larger than the correlation length of the system,
 these subsystems are approximately independent.
 In our case, the correlation length is roughly $10$ or smaller
 (Fig.~\ref{fig:correlation}).
 The number of defects $m$ in the entire system is the sum of defects of
 independent subsystems.
 The mean and variance of $m$
 are then proportional to the system size.
 Moreover, as stated by the central limit theorem,
 $m$ will obey the Gaussian distribution with mean
 $\mu = n \tilde \mu$ and
 variance $\sigma^2 = n {\tilde \sigma}^2$ where $n \equiv \frac{N}{\tilde
 N}$ when $n$ is sufficiently
 large.
 This is also approximately the case for the Neumann boundary condition
 when $L$ is sufficiently larger than the correlation length.

 Because this argument is very general, the Gaussian distribution should be
 obtained for both the periodic and Neumann boundary conditions and
 other models exhibiting spiral chaos when $N$ is sufficiently large.
 In fact, we confirmed it for the
 B\"{a}r model and the CGLE with all the parameter sets we chose and both
 boundary conditions (results not shown).

\begin{figure}
\begin{center}
 \includegraphics[scale=0.25]{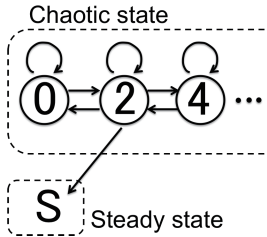}
\end{center}
 \caption{Defect generation-annihilation process for periodic
 boundary condition. The circle with the number $m$ denotes the state
 with $m$ defects. The symbol S denotes the uniform steady state.}
 \label{fig:transition}
\end{figure}

\begin{figure}[b]
\begin{center}
 \includegraphics[scale=0.5]{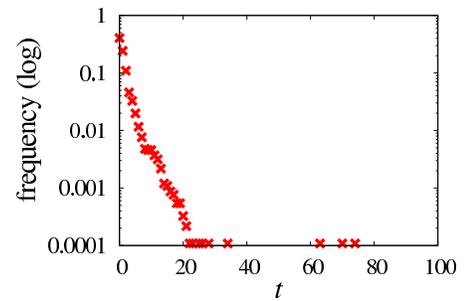}
\end{center}

 \caption{ 
 Normalized histogram of the duration in which the number $m$ of defects
 continues to be zero until $m$ becomes two. Here, the value at $t =
 k \in \mathbb{N}$ denotes the frequency of the duration $k-1 < t \le
 k$.  The B\"{a}r model with $b=0.07,\epsilon=0.09$.
 Defects seldom reemerge for $t > 20$.
}
 \label{fig:defectvanish}
\end{figure}

 \section{system size dependence of lifetime}\label{sec:lifetime}
\begin{figure*}[t]
\begin{minipage}{0.3\hsize}
\begin{center}
 \includegraphics[scale=0.42]{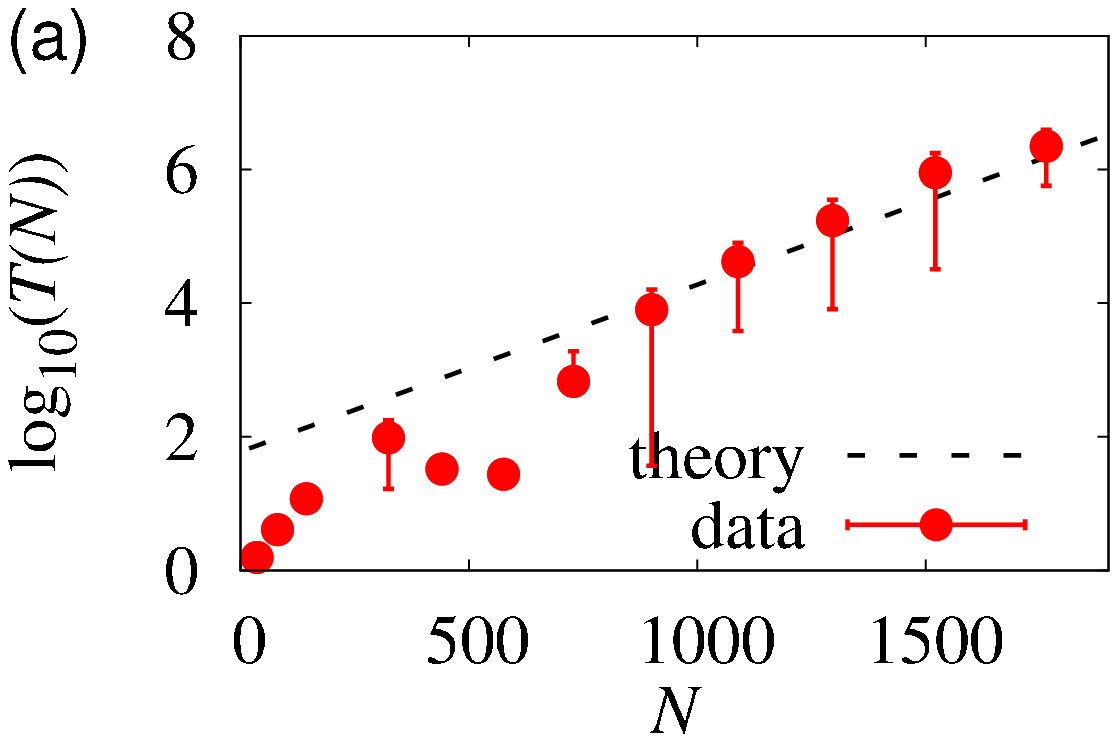}
\end{center}
\end{minipage}
\begin{minipage}{0.3\hsize}
\begin{center}
 \includegraphics[scale=0.42]{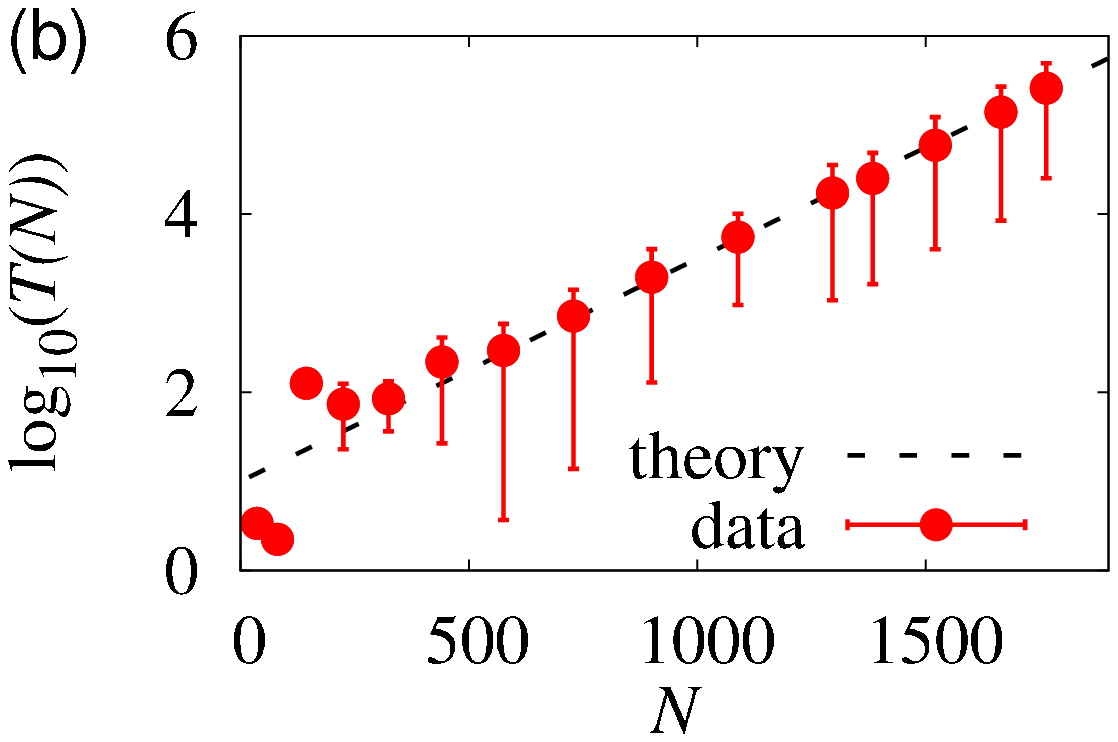}
\end{center}
\end{minipage}
\begin{minipage}{0.3\hsize}
\begin{center}
 \includegraphics[scale=0.42]{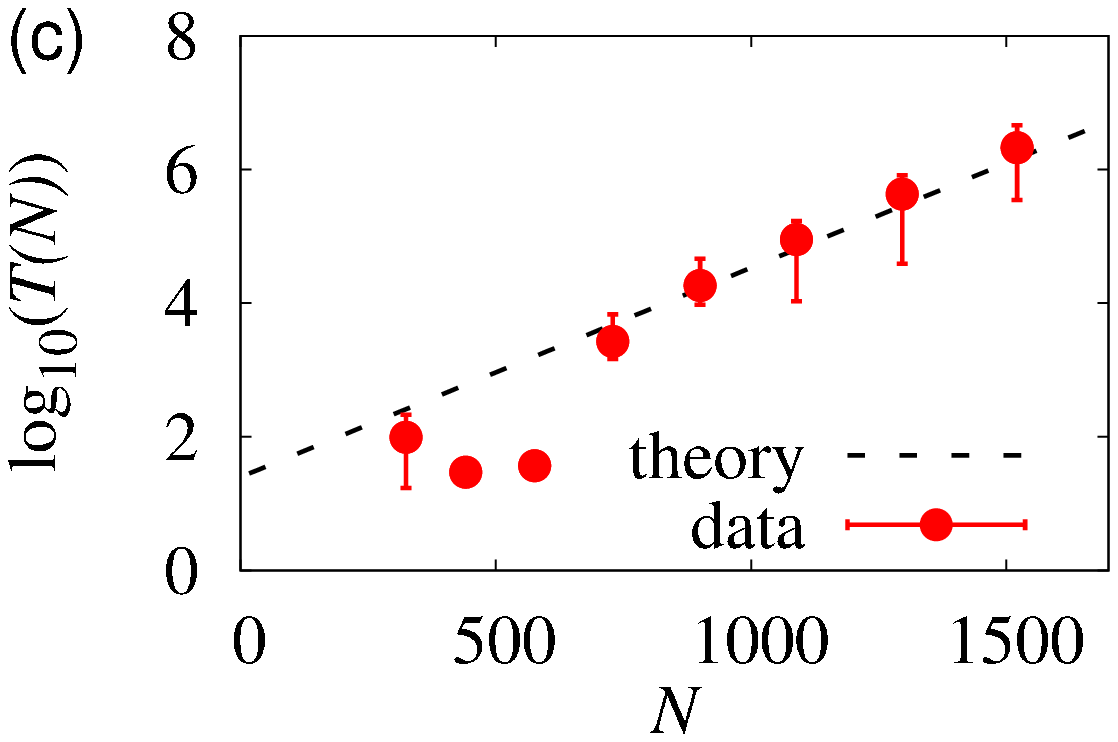}
\end{center}
\end{minipage}
\begin{minipage}{0.3\hsize}
\begin{center}
 \includegraphics[scale=0.42]{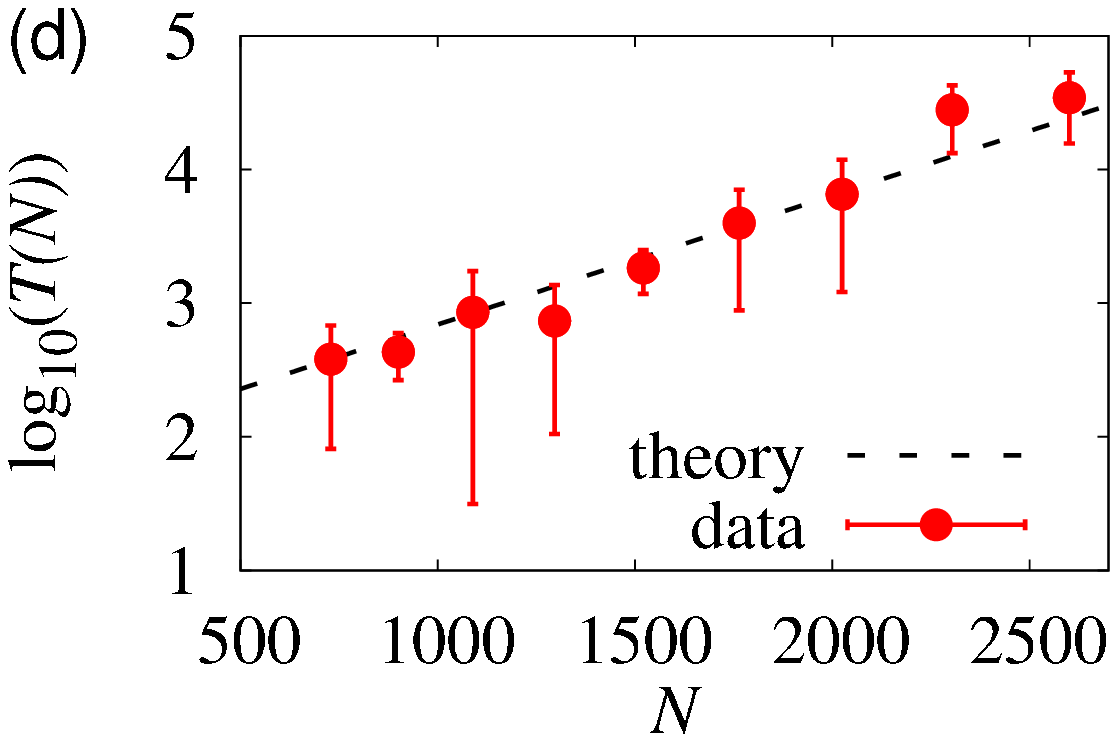}
\end{center}
\end{minipage}
\begin{minipage}{0.3\hsize}
\begin{center}
 \includegraphics[scale=0.42]{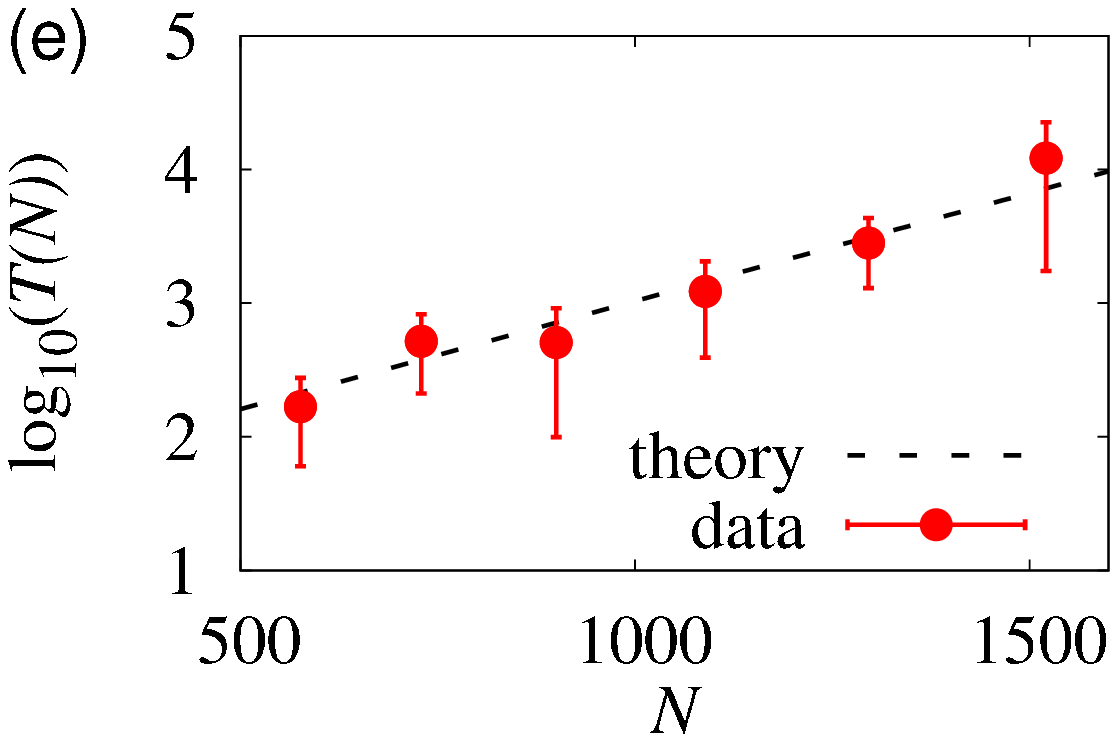}
\end{center}
\end{minipage}
\begin{minipage}{0.3\hsize}
\begin{center}
 \includegraphics[scale=0.42]{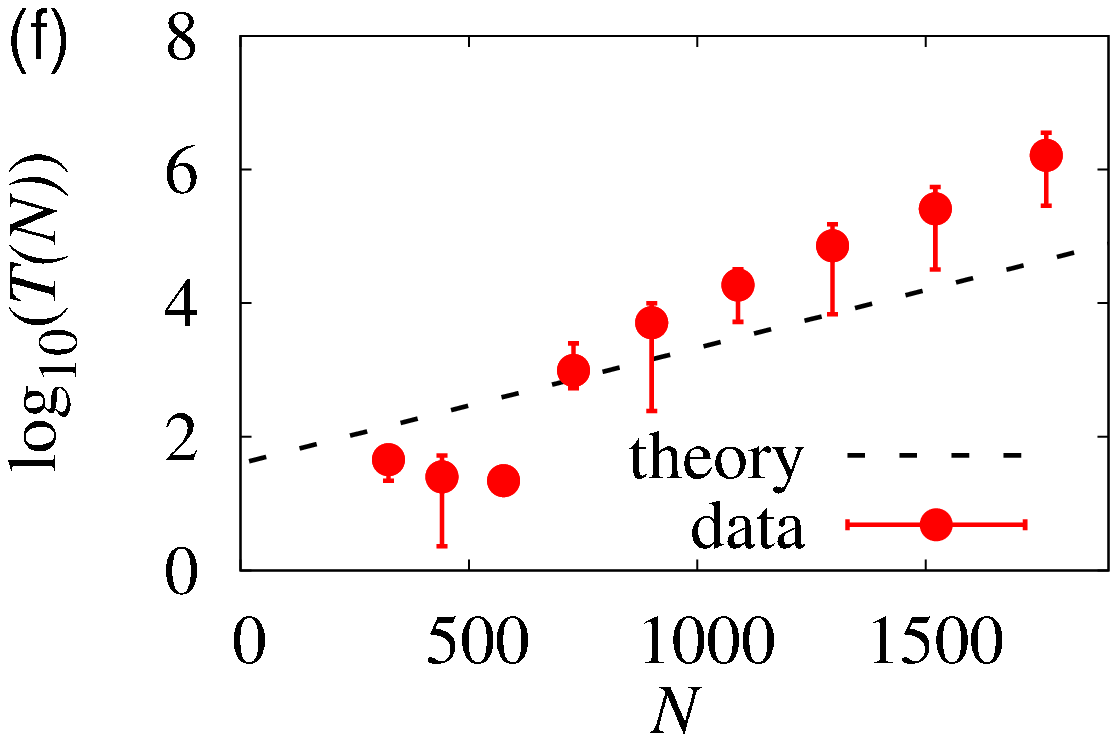}
\end{center}
\end{minipage}
 \caption{Lifetime $T(N)$ (log scale) vs $N$.
 The symbol and the error bar correspond to the average and the
 standard deviation of $T(N)$ for each system size, respectively.
 (a) B\"{a}r model with periodic boundary condition, $b=0.070,~\epsilon=0.080$.
 (b) B\"{a}r model with Neumann boundary condition, $b=0.070,~\epsilon=0.080$.
(c) B\"{a}r model with periodic boundary condition, $b=0.030,~\epsilon=0.075$.
(d) CGLE with periodic boundary condition, $c_1=0.80,~c_2=-1.00$.
(e) CGLE with periodic boundary condition, $c_1=0.50,~c_2=-1.11$.
 (f) B\"{a}r model with periodic boundary condition, $b=0.070,~\epsilon=0.090$.
Note that in the B\"{a}r model with the periodic boundary condition (a,c,f),
the system size dependence of lifetime is obviously not exponential for
small system sizes ($N \le 24^2$). For such small systems, we find 
that transient chaotic states occasionally end up with various complex
 patterns including temporally periodic states with $m \neq 0$.}
 \label{fig:lifetime}
\end{figure*}

 As already mentioned, a previous numerical study reported that
 the lifetime of transient spiral chaos increases exponentially with the system
 size.
 We also numerically confirm it in the following manner.

 In any boundary conditions, all the defects must
 completely vanish before the system settles down to the steady
 state. Here, it should be noted that there is still a chance that a pair of
 defects is generated even from the state with $m=0$
 because of some remaining complex pattern \footnote{See Video 1 and 2 of
 Supplemental Material at [inserted by publisher] for an example of defect regeneration.}.

 Therefore, the transition between the states with different
 numbers of defects $m$ can be illustrated as in
 Fig.~\ref{fig:transition}, where
 the periodic boundary condition is
 assumed for simplicity so that $m$ takes only even numbers, and
 the symbol S denotes the uniform steady state.

 To define the lifetime, we regard
 the system state as the steady state when the duration of
 the state with $m=0$ continues for 100 simulation time, as
 defects hardly reemerge if the state with $m=0$ continues for 20
 simulation times (Fig.~\ref{fig:defectvanish}).
 Under such a numerical setup, we investigate the dependence of the
 lifetime on the system size $N$ (Fig.~\ref{fig:lifetime}), which
 is indeed exponential.

 The expression for the system size dependence of lifetime $T(N)$ can be
 obtained as follows.
 We assume that the process illustrated in
 Fig.~\ref{fig:transition} is Markovian. Starting from some
 initial number $m^*$ of defects, we have a series of defect number at each
 time; e.g., $\{m^*, m^*+2, \cdots, 4, 4, 4, 6, 6, 4, 2, 2, 0, 0, 2, 2, 2, {\rm
 S}\}$, where the symbol ``S'' denotes the event at which $0$ continues for 100 unit
 time (which we regard as the steady state). The lifetime at each trial
 is the length of this series. The expected value of lifetime $T$ is the
 inverse of the probability $\lambda$ to obtain S. Because S is obtained
 only when the previous number is $2$, $\lambda=Z p(2)$ where $p(2)$ is
 the probability to obtain $2$ and $Z$ is the transition rate from the state
 with $m=2$ to the steady state.
 Therefore, the expected lifetime for a given system size $N$ is 
 \begin{equation}
  T(N) = \frac{1}{Z p(2)}.
   \label{T_vs_p0}
 \end{equation}
 For large $N$, the probability distribution
 of the number of defects is well approximated by Eq.~\eqref{gaussian}
 and the mean number $\mu (= \alpha N)$ of defects is large.
 For $m \ll \mu$, we approximately have
 \begin{align}
  p(m) \sim \exp \left( -\frac{\alpha^2}{2\beta}N \right)
  \label{p0}
 \end{align}
 Plugging this into $p(2)$ in Eq.~\eqref{T_vs_p0} and further assuming
 that $Z$ is independent of $N$,
 we finally obtain
 \begin{equation}
  T(N) \sim \exp \left( \frac{\alpha^2}{2\beta}N \right).
   \label{lifetime}
 \end{equation}
 This expression indicates that the lifetime depends exponentially on the
 system size $N$ and its exponent is associated with the density $\alpha$ and
 the magnitude $\beta$ of the fluctuation of the number of defects.
 For the Neumann boundary condition, the steady state can be reached not
 only from the states with $m=2$ by annihilation but
 also from the states with $m=1$ through the absorption of a defect by
 the boundary.
 Therefore, the probability $\lambda$ to
 obtain S is $\lambda = Z_1 p(1) + Z_2 p(2)$ with transition rates $Z_1$
 and $Z_2$. In this case as well, we obtain
 Eq.~\eqref{lifetime} because both $p(1)$ and $p(2)$ can be well
 approximated by Eq.~\eqref{p0} for large $N$.

 Our expression \eqref{lifetime} is numerically verified (Fig.~\ref{fig:lifetime}). 
 The slope given by Eq.~\eqref{lifetime}
 (the dashed lines)
 is in good agreement with
 that obtained numerically in both the B\"{a}r model
 (Fig.~\ref{fig:lifetime} (a--c))
 and the CGLE (Fig.~\ref{fig:lifetime}(d,e)) for large system sizes.

 However,
 we find discrepancy
 for some parameter sets.  In the B\"{a}r model,
 there are considerable deviations for large $\epsilon$ values (e.g.,
 Fig.~\ref{fig:lifetime}(f)). In the CGLE, we also find such cases for
 some parameter sets, e.g., $c_1=0.50$, $c_2=-1.50$ with the periodic boundary
 condition (result not shown). All together, we find that
 the parameter sets for which our theory is valid are typically 
 in the region near the onset of transient chaos \cite{baer93,chate96}.
 A possible reason why our theory fails when the system is far from the
 onset of spiral chaos will be discussed in Sec.~\ref{sec:conclusions}.

 \section{conclusions and discussion}\label{sec:conclusions}
 In the present paper, we have investigated the system size dependence of
 the lifetime of spiral chaos.  We derived an expression for the
 lifetime, given as Eq.~\eqref{lifetime}, utilizing the fact that the
 probability distribution of the number of defects is Gaussian for large
 system sizes. We confirmed that Eq.~\eqref{lifetime} well fits
 numerically obtained $T(N)$ for two different models, the B\"{a}r model
 and the CGLE, with several parameter sets and different boundary
 conditions.

 We emphasize that Eq.~\eqref{lifetime} is useful for the prediction
 of the lifetime of large systems.
 We can precisely estimate $\alpha$ and $\beta$ values
 from observations of the number of defects in a large system.
 The observation of a relatively small system for different initial
 conditions enables us to find the average lifetime $T(N)$.
 Then, using $T(N) \sim \exp\left( \frac{\alpha^2}{2{\beta}}N\right)$, we can estimate the average lifetime for large system sizes.

\begin{figure}[b]
\begin{minipage}{\hsize}
\begin{center}
 \includegraphics[scale=0.4]{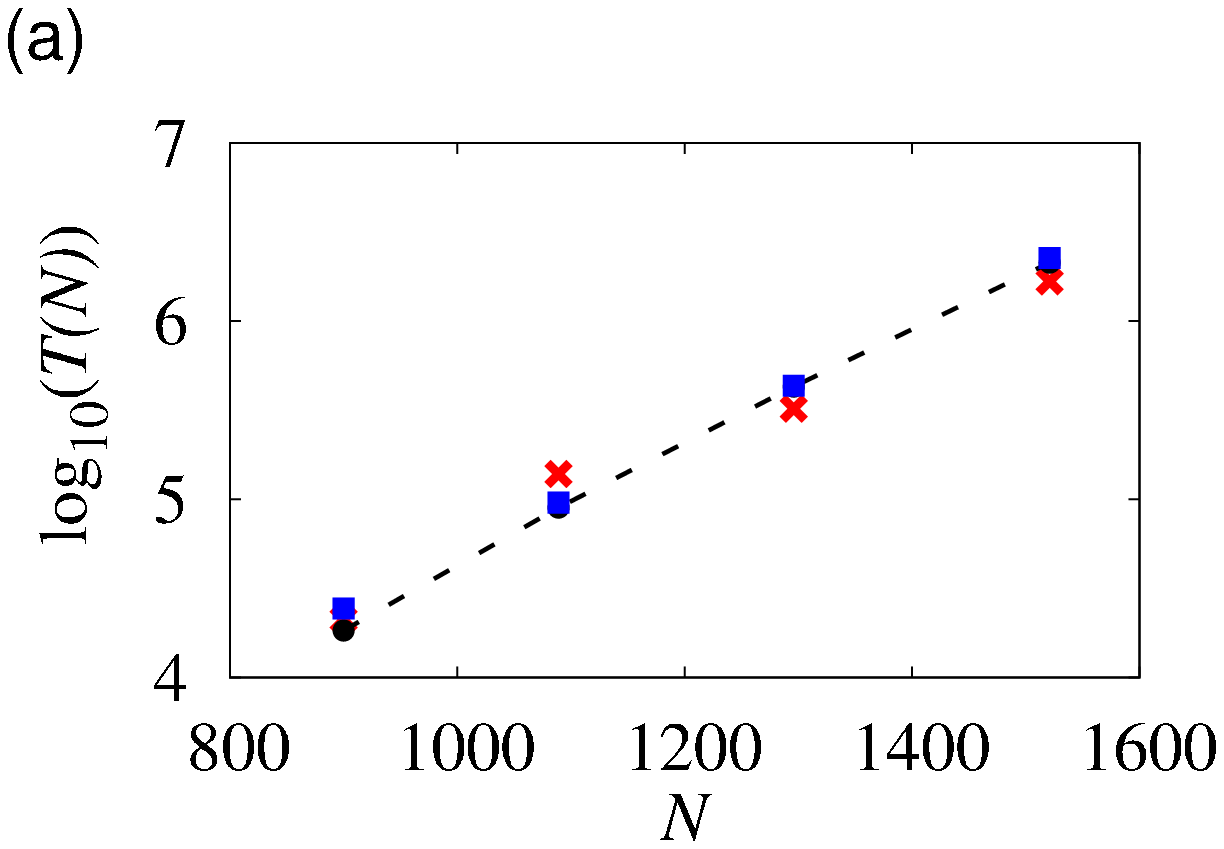}
 \includegraphics[scale=0.4]{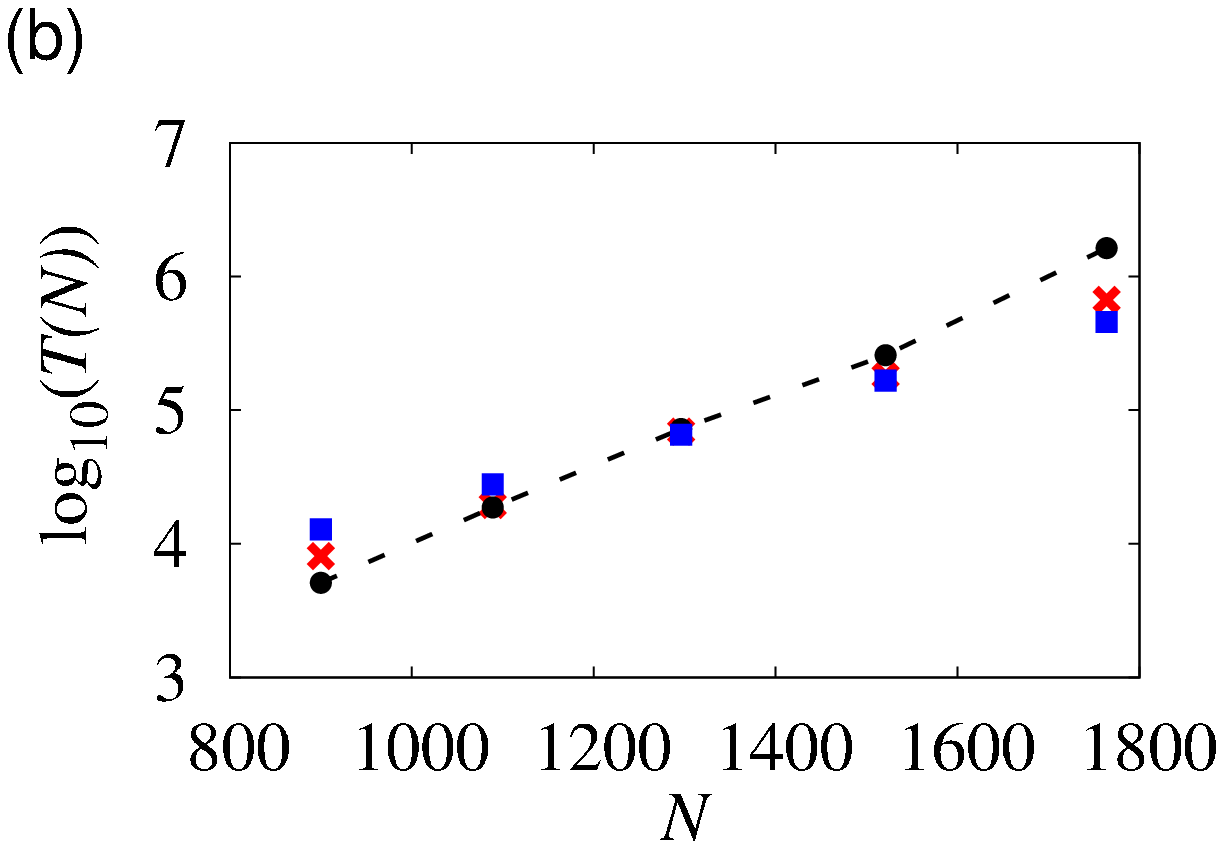}
\end{center}
 \end{minipage}
 \caption{System size dependence of numerically obtained $T$ (symbol $\bullet$ with
 dashed line), $C_1/p(2)$ with numerically obtained $p(2)$
 (symbol $\times$) and $C_2/p(2)$ with $p(2)$ given by Eq.~\eqref{p0}
 (symbol $\Box$), where $C_1$ and $C_2$ are fitting parameters.
 Parameter values for (a) and (b) are the same as those for
 Fig. \ref{fig:lifetime}(c) and Fig. \ref{fig:lifetime}(f), respectively.
}
 \label{fig:p0}
\end{figure}

 We have also found that Eq.~\eqref{lifetime} fails to predict the system
 size dependence for the parameter sets far from the onset of chaos.
 Our theory is based on Eqs.~\eqref{T_vs_p0} and \eqref{p0}.
 We can verify these equations by comparing the system
 size dependences of $T$ and $1/p(2)$ obtained numerically and those
 predicted by Eqs.~\eqref{T_vs_p0} and \eqref{p0}.
 As shown in Fig.~\ref{fig:p0}, whereas both Eqs.~\eqref{T_vs_p0} and \eqref{p0}
 are valid near the onset,
 discrepancy between numerically obtained $T$ and $1/p(2)$ is particularly large
 far from the onset.
 Thus, the assumption in Eq.~\eqref{T_vs_p0} seems to be violated.
 Namely, the transition rate $Z$ from the state with $m=2$ to the
 steady state seems to depend strongly on the system size in such a parameter region.

 The following observation may provide reasoning for it.
 Even when defects completely vanish, 
 some wave pattern may persist for a while. Defect reemergence
 is attributed to such a remaining pattern \cite{Note1}.
 The complexity of wave patterns in the absence of defects might be
 enhanced as the system size increases,
 rendering the system more difficult to settle down in the steady state.
 Indeed, for all parameter sets for which our theory fails, the actual lifetime has stronger
 dependence on the system size than that expected from our theory
 given by Eq.~\eqref{lifetime} with constant $Z$.
 We also observe that 
 meandering of defects and fluctuation in the the number of defects
 seem to be stronger.
 A previous numerical study of the B\"{a}r model
 also indicates that
 the system becomes more strongly chaotic for such parameter sets \cite{strain98}.
 Therefore, it is indeed likely that our system can not be fully
 characterized only by the number of spirals when the system is far from the onset
 of transient chaos.

 \begin{acknowledgments}
 The authors are grateful to Dr. Kei Takayama for
  providing useful information about cardiac arrhythmia.
 They also thank Dr. Sergio Alonso, Dr. Markus B\"{a}r, Dr. Hugues
  Chat\'e, Dr. Hiroyuki Ebata, Dr. Yuki Izumida, Dr. Alexander
  S. Mikhailov and Dr. Mitsusuke Tarama for helpful discussions. 
  This work was supported by the JSPS Core-to-Core Program
 ``Non-equilibrium dynamics of soft-matter and information'' and
  CREST, JST.
 \end{acknowledgments}

\clearpage

\end{document}